\documentstyle[epsfig]{article}
\begin{document}

\title{On the ultra-high energy\\
cosmic ray  horizon}

\author{D. Harari, S. Mollerach and E. Roulet\\ 
CONICET, Centro At\'omico Bariloche,\\
Av. Bustillo 9500, Bariloche, 8400, Argentina}
\maketitle

\begin{abstract}
We compute the ultra-high energy cosmic ray horizon, i.e. the distance up to
which cosmic ray sources may significantly contribute to the fluxes
above a certain threshold on the observed energies. We obtain results both 
for proton and heavy nuclei sources.
\end{abstract}

\section{Introduction}

Soon after the discovery of the cosmic microwave background (CMB)  radiation 
it was realized by Greisen, Zatsepin and Kuzmin \cite{gzk}  that
the fluxes of cosmic ray (CR) protons with energies of order $10^{20}$~eV and 
above would be strongly attenuated over distances of few
tens of Mpc. This is due to the energy losses caused by the photo-pion
production processes in the interactions of the protons with the CMB photons.
Similarly, if CR sources accelerate heavy nuclei, these can photodisintegrate
into lighter ones as they interact with CMB and infrared (IR)
 photons on their journey to
us. In this way the fragments may arrive to the Earth with
 significantly smaller energies than the parent nuclei produced at the sources.
Moreover, both protons and heavy nuclei can further loose energy by pair
production processes, although due to the small inelasticities
involved the typical attenuation length associated to $e^+e^-$ production at
ultra-high energies is large ($\sim 1$~Gpc for protons).

Many works have studied these processes (see
e.g. refs. \cite{st68,hi85,be88,be90,ah90,yo93,ra93,ah94,wa95,be05} 
for the case of protons, and \cite{psb,ep98,st99,kh05,ho06,al05,al06} for nuclei),
and our purpose here is to
make a detailed  analysis to present the results in a way which can be
useful for the study of correlations between the arrival directions of the
highest energy cosmic rays (we will focus here on energies above 50~EeV, 
where 1~${\rm
  EeV}\equiv 10^{18}$~eV) and candidate astronomical sources. Assuming that
the sources are distributed uniformly and have similar absolute cosmic ray 
luminosities, we  will obtain the
distances within which the major part  of the observed events should be
produced (i.e. the horizon for the potential sources) as a function of the
threshold adopted for the energies of the events.

\section{The horizon for protons}

The attenuation length $\lambda$ for the propagation of protons is just
\begin{equation}
\lambda^{-1}\equiv -\frac{1}{E}\frac{{\rm d}E}{{\rm d}x}=
\lambda_{\gamma\pi}^{-1} +\lambda_{ee}^{-1} +\lambda_z^{-1},
\label{dedx}
\end{equation}
receiving contributions from photo-pion ($\lambda_{\gamma\pi}$), pair creation
($\lambda_{ee}$) and redshift ($\lambda_{z}$) losses.

The attenuation length from photo-pion production by the protons in interaction
with the CMB photons is given by (we use hereafter natural units)
\begin{eqnarray}
\lambda_{\gamma\pi}^{-1}&=&\frac{1}{2\gamma^2}\int_0^\infty \frac{{\rm
    d}\epsilon}{\epsilon^2} \frac{{\rm d}n}{{\rm
    d}\epsilon}\int_0^{2\gamma\epsilon} 
{\rm d}\epsilon'\  \epsilon'\eta(\epsilon')\sigma(\epsilon')\cr
&=&-\frac{kT}{2\pi^2\gamma^2}\int_0^\infty
{\rm d}\epsilon'\  \epsilon'\eta(\epsilon')\sigma(\epsilon')\;{\rm ln}\left[
1-\exp(-\epsilon'/2\gamma kT)\right],
\label{gampi.eq}
\end{eqnarray}
where $\gamma=E/m_p$ is the Lorentz factor and the CMB photon density 
 in the observer's rest  frame is
\begin{equation}
\frac{{\rm d}n}{{\rm
    d}\epsilon}=\frac{\epsilon^2}{\pi^2\left(\exp(\epsilon/kT)-1\right)},
\end{equation}
being characterized by a temperature $T=2.73^\circ$K.
 The cross section for pion production in terms
of the photon energy $\epsilon'$ in the proton's rest frame is
$\sigma(\epsilon')$ and the inelasticity factor 
$\eta(\epsilon')$ is the average fraction of the proton energy lost in the
process. We use for the product $\sigma\eta$ a fit to the average values
given in \cite{ra93}, which includes both single pion production near 
the $\Delta$ resonance and  multi-pion production at higher energies.

This attenuation length, as well as that due to pair production obtained
following \cite{bl70,ch92}, are depicted in fig.~1 as a function of the proton
energy. The redshift losses, characterized by $\lambda_z\sim c/H_0\simeq
4$~Gpc,  are of
no relevance for the energies considered in the present work, for which the
 effects due to the interactions with the photon background are the dominant
 ones.

\begin{figure}[ht]
\centerline{{\epsfig{width=2.5in,file=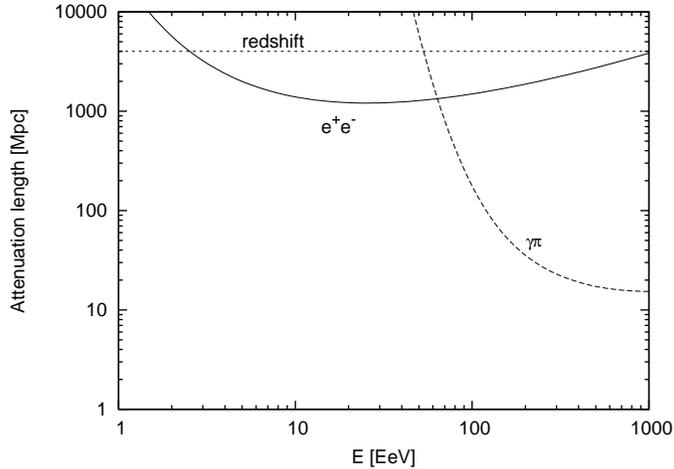,angle=-90}}}
\caption{Proton attenuation length vs. energy.} 
 \label{attlp.fig}
\end{figure}

A direct way to compute the CR horizon is to follow the trajectories of many
individual protons with initial energies distributed according to a given
input spectrum, here taken as a power law one, d$N/{\rm d}E\propto
E^{-\alpha}$ (we will show results for $\alpha=2.2$ and 2.7). We then obtain
the attenuation factor $A(E_{th},x)$, which is defined as the fraction of the
particles which originally had an energy above a threshold value $E_{th}$ 
that still have an energy above that threshold after traversing a distance
$x$.
Assuming that the sources are uniformly distributed and have similar
intrinsic CR luminosities and spectra, one finds that the fraction of the
events observed above a given energy threshold which originated in sources 
farther than a distance  $D$ is just
\begin{equation}
F(E_{th},D)={\int_D^\infty {\rm d}x\ A(E_{th},x) \over 
\int_0^\infty {\rm d}x\ A(E_{th},x)}.
\label{frac.eq}
\end{equation}
Since we will be interested in threshold energies above 50~EeV,
cosmological effects due to the Universe's expansion or to source evolution
are negligible.
This implies that the effects of the inverse square distance reduction
of the fluxes from each source and the increase in the number of sources with
distance compensate each other, leaving just the simple 
integrals in eq.~(\ref{frac.eq}).

\begin{figure}[ht]
\centerline{{\epsfig{width=3.in,file=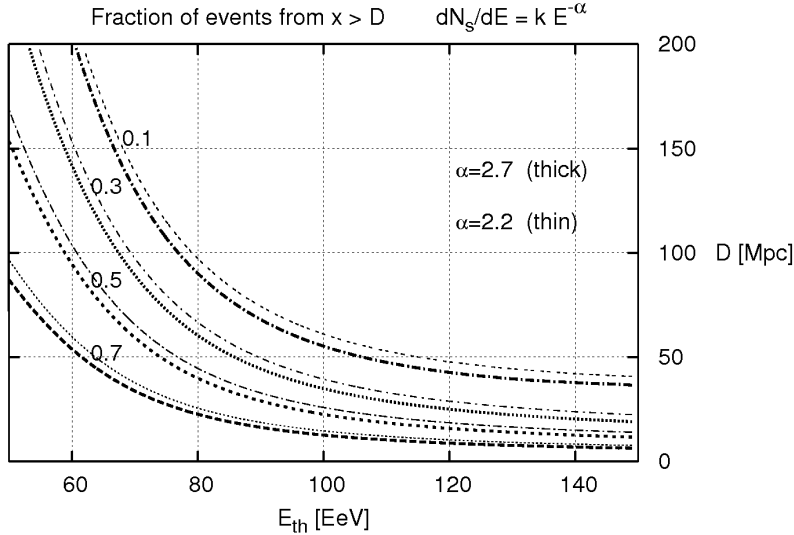,angle=-90}}}
\caption{Considering the events arriving with energies above $E_{th}$, the
  contours indicate the fraction originating  from
  distances larger than $D$, assuming a uniform distribution of proton
sources with  similar intensity.} 
 \label{fracp.fig}
\end{figure}

The fraction $F$ is depicted in fig.~2. We see that the horizon for
protons, which may be taken as the distance for which this fraction reaches
e.g. 10\%, is relatively close on cosmological grounds for all the energies
considered.  For instance, for $E_{th}=80$~EeV  one has that 
90\% of the events should have
been produced at distances not farther than $\sim 90$~Mpc\footnote{It is not
  straight-forward to guess the results in fig.~2 from those in fig.~1, mainly
  because fig.~2 is in terms of the threshold energy measured on Earth, so
  that the CRs involved have energies above the thresholds and moreover their
  energies were even larger at the sources.}. The sensitivity to
the assumed source spectral index $\alpha$ is not large, although as expected
the horizon increases for harder spectra since above a given threshold 
the fraction of higher energy events, which are more penetrating,
becomes larger in this case. Let us also mention that the pair production
losses have an impact on the results only for $E_{th}<80$~EeV, and they become
indeed quite important for $E_{th}<60$~EeV.

Note that since for a power law spectrum the number of events above a given
threshold is proportional to $E_{th}^{-\alpha+1}$, one has simply that
$A(E_{th},x)=\left(E_{th}/E'_{th}(x)\right)^{\alpha-1}$, where $E'_{th}(x)$ is
the initial energy that a CR proton should have in order to arrive with an
average 
energy $E_{th}$ after traversing a distance $x$. The energies $E'_{th}$ can be
easily computed by solving the evolution equation, d$E/{\rm d}x=-E/\lambda$,
backwards in time. We checked that this simpler 
procedure leads indeed to the same
results as the Monte Carlo computations discussed before, but it has the
drawback that it cannot be generalized to the case of heavy nuclei, since for
these last it is not possible to know in advance the final average mass of the
fragment after the nucleus has propagated a given distance.

Let us mention that we are using the so-called continuous energy loss
approximation to evolve the CR energies. This does not account for the
fluctuations originating from the stochastic nature of the pion production
processes. Although these fluctuations could be relevant for the study of
detailed features in the differential CR  spectrum (such as the `bump'
preceding the GZK cutoff), they have only a minor impact on the quantities we
compute, which are based on the integral spectra above specified thresholds. 

\section{The horizon for nuclei}

\begin{figure}[ht]
\centerline{{\epsfig{width=2.5in,file=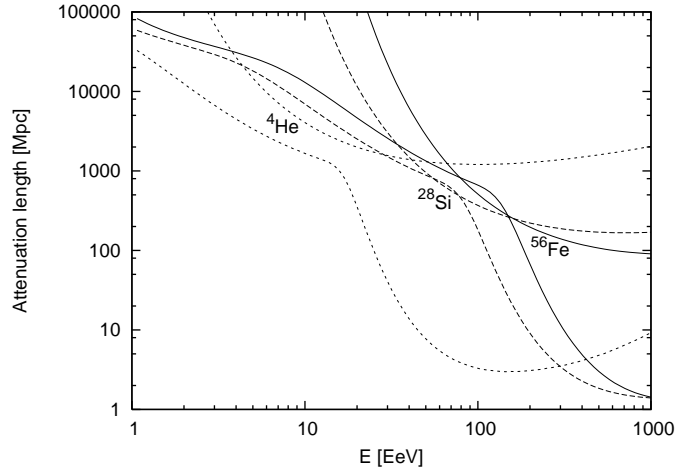,angle=-90}}}
\caption{Attenuation lengths vs. energy for different nuclei. Lower curves are
due to photo-disintegration processes and upper curves to pair 
production processes.} 
 \label{attnuclei.fig}
\end{figure}

The rate at which a nucleus of mass number $A$ photo-disintegrates with the
emission of $i$ nucleons is given by
\begin{equation}
R_{A,i}=\frac{1}{2\gamma^2}\int_0^\infty \frac{{\rm
    d}\epsilon}{\epsilon^2} \frac{{\rm d}n}{{\rm
    d}\epsilon}\int_0^{2\gamma\epsilon} 
{\rm d}\epsilon'\  \epsilon'\sigma_{A,i}(\epsilon'),
\end{equation}
where now $\gamma=E/Am_p$.
The relevant photon densities in this case are the CMB and the IR
backgrounds. For this last we  use the estimates obtained in \cite{ma01}.
The cross sections $\sigma_{A,i}$ 
for the different processes in which a nucleus of mass number $A$ emits $i$
nucleons were parametrised in
\cite{psb}, and we also use the updated energy threshold values presented in
\cite{st99}. 
To describe the average energy loss of a given nucleus it is convenient to
introduce the effective rate \cite{psb}
\begin{equation}
R_{A,eff}\equiv \sum_i i R_{A,i}.
\end{equation}
In terms of this rate one has
\begin{equation}
\frac{{\rm d}A}{{\rm d}x}=-R_{A,eff}.
\label{dadx}
\end{equation}
The attenuation length for photodisintegration is then
\begin{equation}
\lambda_{\gamma d}^{-1}=-\frac{1}{A}\frac{{\rm d}A}{{\rm d}x}=\frac{1}{A}
R_{A,eff}.
\end{equation}
The quantities $\lambda_{\gamma d}$, together with the attenuation lengths to
pair creation, are depicted in fig.~3 for different nuclei: $^{56}$Fe,
$^{28}$Si and $^4$He. It is apparent that pair creation losses become
subdominant for all energies in the case of nuclei lighter than Si, while
for instance in the case of Fe they are dominant for $E\simeq 100$~EeV.
 On the other hand, for $A>20$ one has that to a very good approximation 
\begin{equation}
\lambda_{\gamma d}(A,E)\simeq \lambda_{\gamma d}(56,56\,E/A).
\label{scaling}
\end{equation}
For light nuclei the attenuation lengths are in general 
somewhat larger than what would be
implied by the above relation, due in particular to the less important role
played by multi-nucleon emission in these cases and 
the narrower giant dipole resonances of lighter nuclei\footnote{Other 
peculiarities
  affect some nuclei, such as is the case for Beryllium, which disintegrates
  via ${\rm Be}\to 2\,{\rm He+p}$.}. 
Anyway the general scaling of the results when
different nuclei are considered can be understood from eq.~(\ref{scaling}).

\begin{figure}[ht]
\centerline{{\epsfig{width=3.in,file=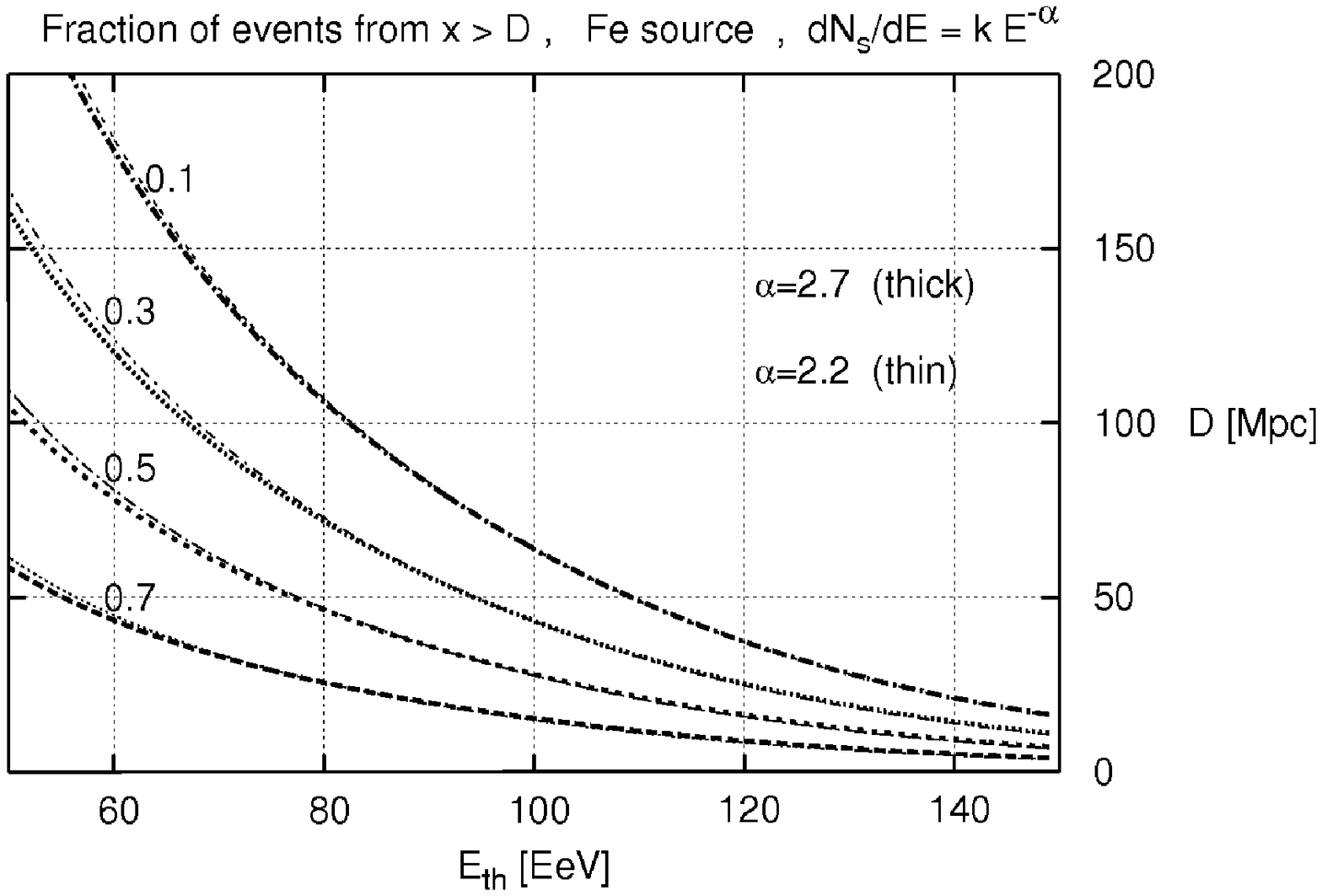,angle=-90}}}
\caption{Considering the events arriving with energies above $E_{th}$, the
  contours indicate the fraction originating  from
  distances larger than $D$,  assuming a uniform distribution of Fe 
sources with  similar intensities.} 
 \label{fracnuc.fig}
\end{figure}

\begin{figure}
\centerline{{\epsfig{width=3.in,file=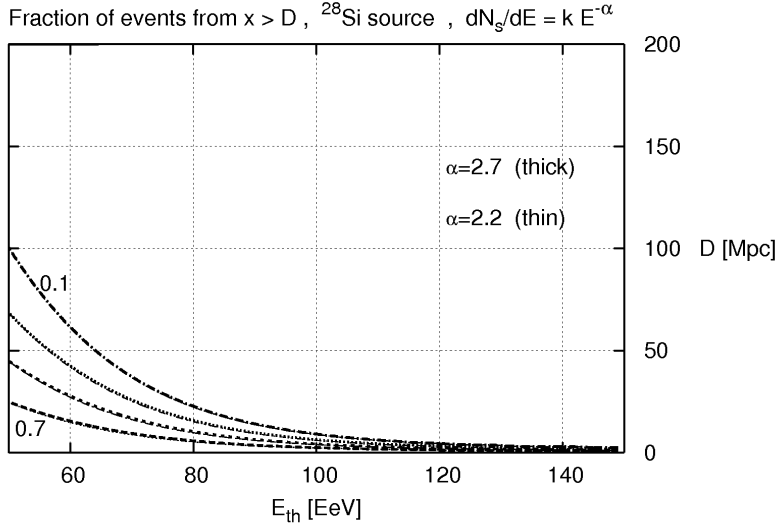,angle=-90}}}
\caption{Considering the events arriving with energies above $E_{th}$, the
  contours indicate the fraction originating  from
  distances larger than $D$, assuming a uniform distribution of Si 
sources with  similar intensities.} 
 \label{fracsi.fig}
\end{figure}

In fig.~4 we depict the horizons for sources producing fluxes of Fe nuclei,
obtained by following a large number of CRs with energies distributed 
according to the assumed spectra and obtaining the attenuation factors and
associated fractions, as done before for the proton sources (fig.~2). 
For the evolution of the mass number we use eq.~(\ref{dadx})\footnote{For a
  given mass number we take the charge as the smaller one corresponding to
  a stable isotope.}, while for the
energies we account also for pair creation losses.
In the
Fe source case the horizons result smaller than for proton sources and they
become even smaller for lighter nuclei, for which the attenuation lengths 
become shorter at the
energies we are considering (fig.~3). This is illustrated in fig.~5,
 corresponding to the case of $^{28}$Si sources. 
It is clear that in a realistic situation the actual fractions will depend on
the details of the non-homogeneous local source distribution and on the source
intensities, but the limiting distances from which CRs can arrive above a
given threshold are quite insensitive to these details.

\begin{figure}[ht]
\centerline{{\epsfig{width=2.in,file=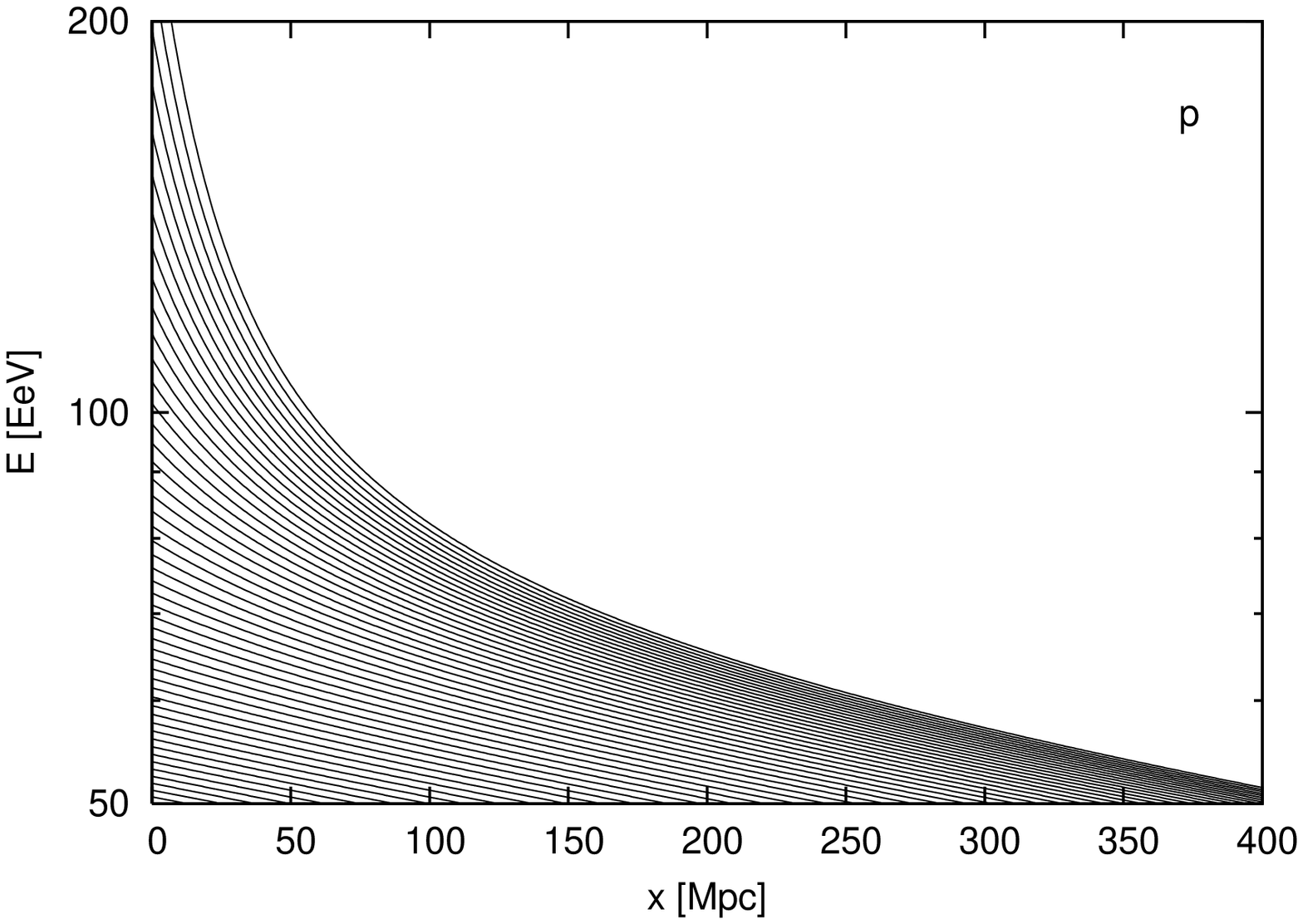,angle=-90}}}
\centerline{{\epsfig{width=2.in,file=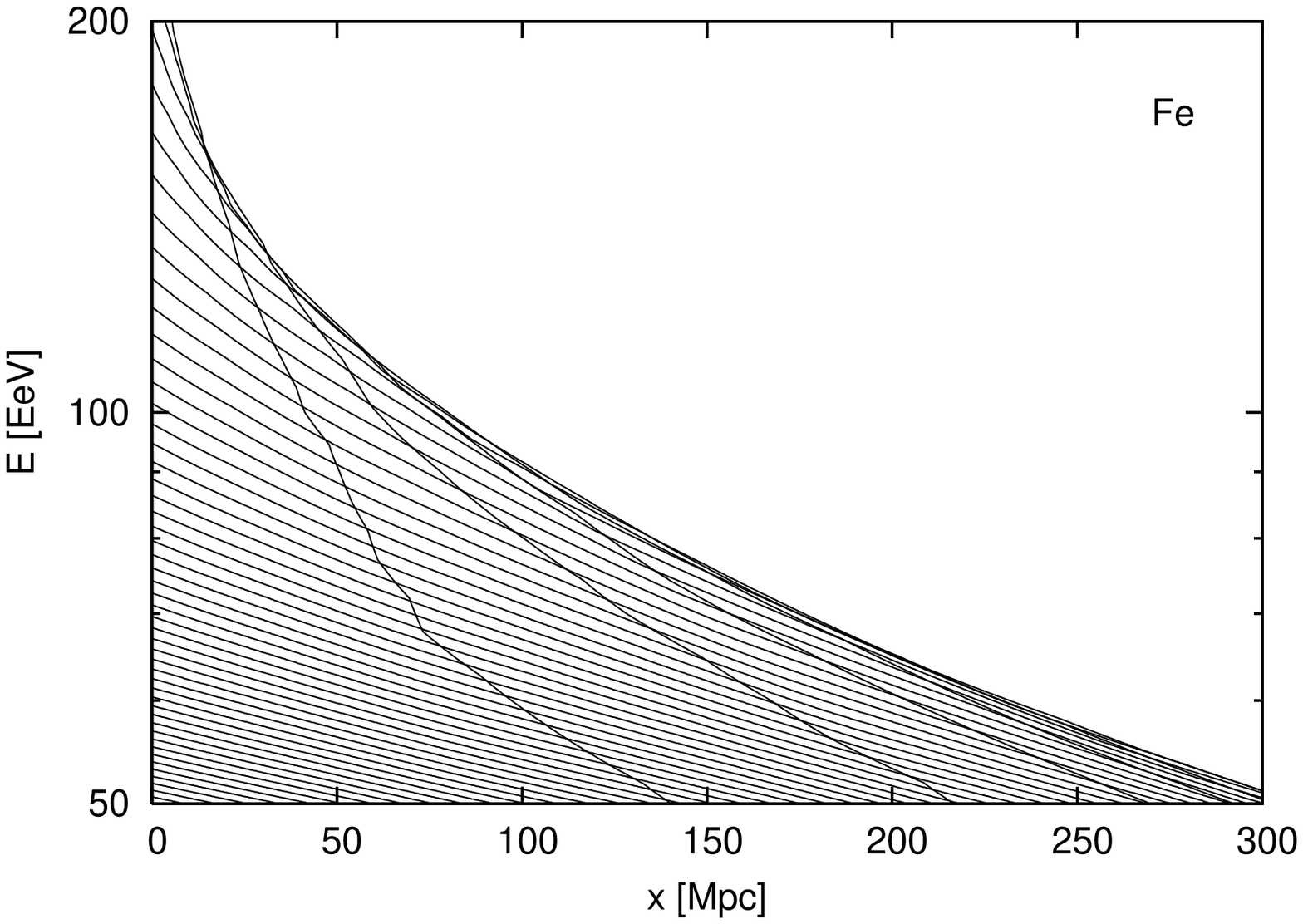,angle=-90}}
{\epsfig{width=2.in,file=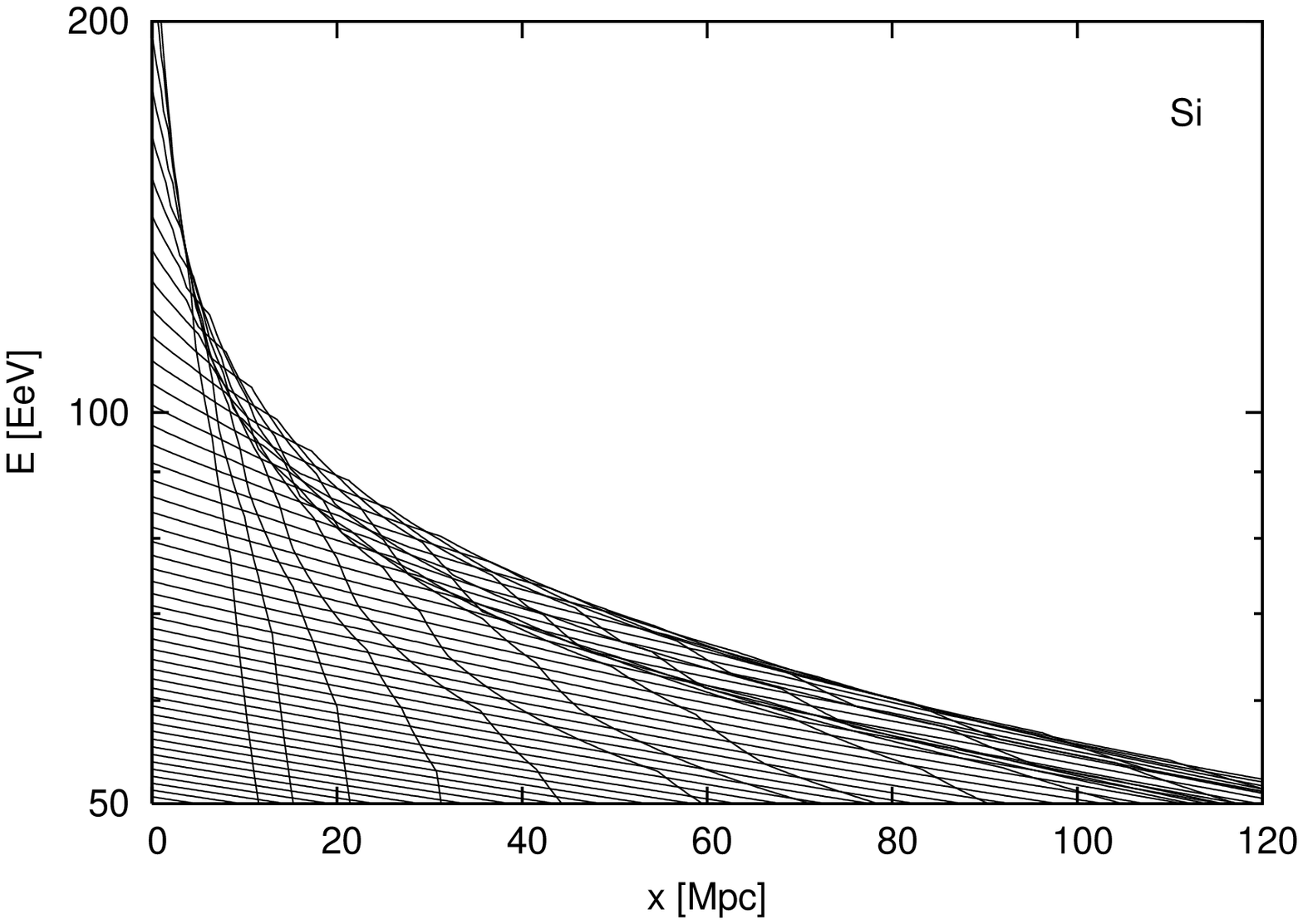,angle=-90}}}
\caption{Energies as a function of distance to the source for a sample 
of initial energies. Top figure is for protons and bottom
ones for Fe and  Si sources (note the different horizontal ranges).}  
 \label{evsx.fig}
\end{figure}

To better understand these results we show in figs.~\ref{evsx.fig} the
energies as a function of distance for a set of 50 CRs with initial 
energies sampling an $E^{-2.7}$ spectrum for the three cases discussed (p, Fe
and Si). The results in figs.~2, 4 and 5 can then be interpreted
naturally. 

An important difference between the proton and nuclei 
results is related  to the fact
that when a nuclear species is above the threshold of the giant dipole
resonance for a typical CMB photon, also the fragments from the
photodisintegration will be above threshold (neglecting the subdominant energy
losses due to pair creation) and hence the disintegration
continues efficiently until the nucleus is completely disintegrated. On the
contrary, nucleons which are originally above the threshold for photo-pion
production and loose energy by pion emission may soon end up below threshold
and hence continue their travel without further losses due to pion production.
 This leads to the well
known pile-up in the proton spectrum, which is well seen in the top panel of 
fig.~(\ref{evsx.fig}), 
 a feature which is much less pronounced
 in the case of nuclei. This effect is also responsible for
the smaller sensitivity
of the results in figs. 4 and 5 to the adopted spectral index.

\begin{figure}[ht]
\centerline{{\epsfig{width=2.5in,file=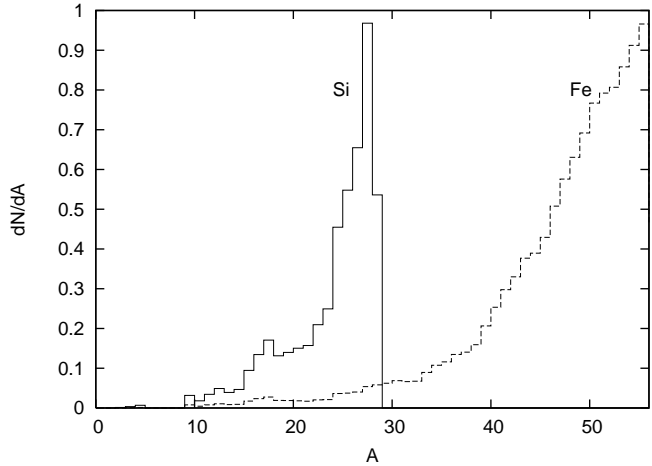,angle=-90}}}
\caption{Distribution of masses of the fragments arriving to the Earth with
$E>60$~EeV in a
  scenario with uniformly distributed Fe (dashed) and Si (solid)
sources with spectral slope
  $\alpha=2.7$ (arbitrary units).}  
 \label{dnda.fig}
\end{figure}

It is interesting to note that even if  these scenarios
adopt a single nuclear mass at the sources,  the
distribution of the masses of the CRs arriving to the Earth 
is quite wide, with contributions from many different nuclear species.
This is exemplified in fig.~\ref{dnda.fig}, 
which displays the relative distribution of
mass fragments arriving to the Earth above a threshold energy of 60~EeV for 
scenarios with Fe (dashed histogram) and Si (solid histogram)
 sources, with  spectral index $\alpha =2.7$ (the results are not very
 sensitive to the value of $\alpha$ adopted).
 
The flux of secondary nucleons produced in the various photodisintegration
processes is generally quite suppressed, since these nucleons will be emitted
with 
energies $E/A$ if the CR started with energy $E$ and mass $A$. Hence, 
the secondary nucleons can contribute
non-negligibly to the CR fluxes only in
the case of very hard spectra ($\alpha\simeq 2$) and high source cutoffs
($E_{max}\gg AE_{th}$). For simplicity, we have just assumed in our
computation that the maximum energy attained in the sources is
$E_{max}=AE_{th}$ (which is bigger than $10^{21}$~eV in the examples 
considered),
 and in this case the secondary nucleons are always below threshold. 

It is also worth mentioning  that, as is seen from fig.~3, heavy
nuclei produced with energies $E>A\times 10$~EeV have attenuation lengths
of only a few Mpc. Hence, these nuclei will be
essentially completely disintegrated after traveling $\sim
10$~Mpc. 
Since each nucleus of initial mass $A$ and energy $E$ will lead to $A$
nucleons with energies $E/A$, the number of nucleons above an energy threshold 
$E_0$ will be in this case $N(>E_0)\simeq AN_A(>AE_0)$, with
$N_A(>E)\propto E^{1-\alpha}$ being the original integral source spectra
 (assuming here that the power
law flux has no upper cutoff).
We hence learn that the
sources of  nuclei  farther than $ 10$~Mpc will 
produce a flux of secondary nucleons above 10~EeV similar to the 
 source flux but  suppressed by a factor $A^{2-\alpha}$, and extending up to
 energies $E_{max}/A$.  The
attenuation of these secondary  nucleons caused by photo-pion and pair creation
losses may then be studied  as done in section~2.

Let us finally mention that both pair production losses and
photo-disintegrations with the IR background have a significant impact,
especially for the lower thresholds considered, on the results for Fe nuclei
(fig.~4), while in the case of Si sources (fig.~5) 
their effect is less important.

\section{Conclusions}
As a summary, we have shown that the sources of the vast majority of the CR
events observed with energies above 50~EeV should be at relatively close
distances (under the plausible assumption that CRs consist of ordinary
nucleons or nuclei), as is illustrated in a quantitative way in
figs.~2, 4 and 5\footnote{Some
  particular values of the results in fig.~2 can be read off from plots in
  previous works \cite{wa95,wa96,cu05}. To our knowledge the results for nuclei
  are not available in the literature.}. These results should be relevant in
order to restrict  the distances to the candidate
astrophysical objects  that could act as sources for the observed ultra-high
energy cosmic rays. 

\section*{Acknowledgments}

We are grateful to ANPCyT (grants PICT 13562-03 and 10881-03)
and CONICET (grant PIP 5231) for financial support.

\end{document}